\documentclass[twocolumn,showpacs,preprintnumbers,amsmath,amssymb]{revtex4}
\usepackage[dvips]{graphics}
\usepackage{color}

\newcommand{\beq}{\begin{equation}}
\newcommand{\eeq}{\end{equation}}
\newcommand{\bma}{\begin{math}}
\newcommand{\ema}{\end{math}}
\newcommand{\beqa}{\begin{eqnarray}}
\newcommand{\eeqa}{\end{eqnarray}}
\def\opone{\le\textbf{}\textbf{}avevmode\hbox{\small1\kern-3.8pt\normalsize1}}

\newcommand{\be}[1]{     \begin{eqnarray} \mbox{$\label{#1}$}   }

\newcommand{\ee}{\end{eqnarray}}

\begin{document}

\title{Tao-Thouless Revisited}

\author{T.H. Hansson}

\author{A. Karlhede}

\affiliation{Department of Physics,
Stockholm University \\ AlbaNova University Center\\ SE-106 91 Stockholm,
Sweden}

\date{\today}

\begin{abstract}
It is now established that the state proposed by Tao and Thouless for the fractional quantum Hall effect in 1983, 
shortly after Laughlin's work, is the ground state in the so-called Tao-Thouless limit, and that it is adiabatically 
connected to the Laughlin state. We review the interesting history of the Tao-Thouless state, and its generalizations, 
and discuss its relevance and shortcomings. 
In particular, we calculate the exclusion statistics of quasiparticles in the Tao-Thouless limit and point out a principal difficulty which 
prevents  calculation of the fractional exchange statistics in this limit. 
\end{abstract}
\pacs{73.43.Cd, 71.10.Pm, 75.10.Pq}

\maketitle

\section{The Tao-Thouless state}\label{TT}

Shortly after Laughlin's article explaining the fractional quantum Hall (QH) effect at filling factors $\nu=1/(2k+1)$ appeared in 1983 \cite{Laughlin83}, 
Tao and Thouless proposed an alternative homogeneous ground state, with a gap to excitations \cite{TaoThouless83}.
The Tao-Thouless (TT) state is the single Slater determinant state where, on a cylinder in Landau gauge for $\nu=1/3$, every third of the one-electron 
states are occupied:  $001001001001\dots$, {\it ie} the state with unit cell $001$.

The Tao-Thouless state was soon abandoned as a serious candidate for the fractional quantum Hall ground state. This was based on the small 
overlap of the state with the exact ground state for small systems as well as on Thouless's analysis 
of the exchange hole in the pair correlation function  \cite{Thouless85}. Thouless concluded that the Laughlin and the TT-states 
were completely different and that the evidence was strongly against the latter as a correct description of the quantum Hall effect---he wrote 
''I think the time has come to abandon the Tao-Thouless theory'' \cite{Thouless85}. However, in the acknowledgements of the 
same article he wrote ''I have benefited from numerous discussions with  Dr. R. Tao, but have no reason to believe that he 
accepts either  the arguments presented here or the conclusion drawn from them.''.   

It is now established that the Tao-Thouless state is in fact relevant for the description of interacting two-dimensional electrons in a magnetic field. 
If the quantum Hall problem is studied on a cylinder as a 
function of its radius, then the Tao-Thouless state is the exact ground state as the radius goes to zero for generic electron-electron interactions. 
Moreover, the TT-state has the same qualitative properties as the Laughlin state, such as a gap to excitations, and quasiparticles with charge 
$\pm e/(2k+1)$---these are domain walls between degenerate ground states. For example, the underlined segments  in 
$001001\underline{000}1001001\dots$ and $00100\underline{101}001001\dots$,  which consist of three consequitive sites with zero and two electrons respectively, 
are domain walls between the degenerate ground states and contain the quasihole 0 with charge $e/3$ and the quasielectron 01 with charge $-e/3$. Note that the quasihole is obtained by inserting  an additional empty site  0  in the $1/3$-state ground state with unit cell 001, whereas the quasielectron is obtained by inserting 01 (or, equivalently by removing 0). In fact, the Laughlin and TT states are adiabatically connected.  
Moreover, this generalizes to other filling factors: a generalized Tao-Thouless state is 
the ground state at $\nu = p/q$ on the thin cylinder and there is, for $q$ odd, evidence that it is the limit of a possible abelian bulk quantum Hall state.\footnote{Bulk here refers to the standard two-dimensional case realized in an experiment. Our Tao-Thouless states differ from the generalizations presented by Tao \cite{Tao84}.}

This state of affairs has become clear only recently, but in hindsight many of the basic facts were discovered, or argued for, by various 
people over the years. To a large extent this work passed unnoticed. We will here briefly review the important precursors  to the 
recent work that we are aware of. 

\section{History of the Tao-Thouless state}
In 1983, Anderson observed that Laughlin's wave function has a broken discrete symmetry and that the
quasiparticles are domain walls between the degenerate ground
states \cite{anderson}. Furthermore he noted that the TT-state is
non-orthogonal to the Laughlin state and suggested that it can be
thought of as a parent state that develops into the Laughlin wave
function, without a phase transition, as the electron-electron
interaction is turned on. In 1984, Su concluded, based on exact
diagonalization of small systems on the torus, that the QH state at
$\nu=p/q$ is $q$-fold degenerate and that the quasiparticles with charge $\pm e/q$ are domain
walls between these degenerate ground states \cite{su}. 

In 1994, Rezayi and Haldane studied the Laughlin wave function on a
cylinder as a function of its radius \cite{Haldane94}. The method they employed was to use that the Laughlin state is the exact ground state, 
with a gap to excitations, for an ultra-short range interaction \cite{haldanebook,shortrange}. This had previously been 
used on the plane and the sphere, but since the result depends only on how the wave function vanishes when two electrons 
approach each other, it holds also on a cylinder of {\it any} radius. Rezayi and Haldane found that the Laughlin state  approaches
the Tao-Thouless state when the radius of the cylinder goes to zero, and that there is a rapid crossover to a virtually homogeneous state when 
the radius grows.\footnote{Density modulations will remain as long as the radius is finite as noted earlier by Haldane \cite{haldanebook}.}
Since the Laughlin state is the exact ground state {\it and} there is a gap to excitations 
for any radius of the cylinder, it follows that the Tao-Thouless state 
is continuously connected to the Laughlin state, {\it ie} there is no phase transition 
separating them. However, this conclusion was never explicitly drawn in Ref. \onlinecite{Haldane94}, although the basic facts 
are there.
The authors of Ref. \onlinecite{bk1} were informed about this work  
when they were investigating the half-filled Landau level on the thin cylinder \cite{Moessner}. The correct interpretation---that there is no phase transition when going from the Tao-Thouless state to the Laughlin state---is given in Ref. \onlinecite{bk1}, where it is used as circumstantial evidence for there being a continuous connection also for the half-filled Landau level.

\section{Adiabatic continuity}

Adiabatic continuity normally refers to two states being connected as a parameter in the Hamiltonian is changed, whereas in the case at hand the 
radius of the cylinder on which the electrons live changes. However, by writing the Hamiltonian for interacting electrons in a single Landau level in the occupation
number basis, where the creation operators create electrons in the single-particle states in Landau gauge, one sees that only the matrix elements change as the radius of the 
space changes. Thus, rather than viewing the process as taking place on a cylinder as the radius changes, one can view it as taking place on a
cylinder with fixed, possibly infinitely large, radius and changing matrix elements \cite{bk3}. Thus we have the standard setting for discussing adiabatic continuity. Note that from this 
''adiabatic point of view'' the Tao-Thouless state is spatially homogeneous if the radius of the cylinder is chosen to be infinite; this is of course what Tao and Thouless 
originally had in mind in Ref. \onlinecite{TaoThouless83}.  That the limit can be thought of as being taken in a 
fixed space is a reason why we prefer to use the name Tao-Thouless limit rather than thin cylinder, or thin torus, limit. In addition, it gives proper reference to the original work.

A common objection to the claim that  the Laughlin and TT-states are qualitatively similar in the sense that they 
describe the same state of matter, is that the former is homogeneous whereas the latter has a charge density wave structure. 
We have seen that if the adiabatic point of view, with an infinite radius, is taken, then the TT-state is also homogeneous. However, on a cylinder with finite radius the 
TT-state has a crystalline structure but  this
does not imply that the states 
are qualitatively different. First of all, as noted by Haldane \cite{haldanebook}, 
on a torus, and also on the cylinder, there is no perfectly translationally invariant state for a finite system---the Laughlin state, as well as any other state, on the cylinder may become fully homogenous only as the radius goes 
to infinity. Secondly, although the TT-state has a crystalline structure it is not an ordinary crystal as
there are no gapless excitations, no phonons. The reason for this is that the Hamiltonian for a charged particle on a cylinder in a 
homogeneous magnetic field is only invariant under discrete spatial translations, the continuous 
spatial symmetry is explicitly broken, see {\it eg} Ref. \onlinecite{Jansen}.  In a magnetic field, ordinary translations 
are replaced by  magnetic translations as symmetry operations and the Laughlin and the TT states have the same magnetic quantum numbers on the torus.

Sometimes the TT-state is referred 
to as a Wigner crystal. In our opinion this is an unfortunate terminology, as is the use of the name crystal in general in this context, 
since this normally refers to a state with gapless excitations.  The Laughlin and the TT states have the same qualitative properties: they 
have a gap to excitations, quasiparticles with fractional charge $\pm e/(2k+1)$ and the same magnetic quantum numbers when put on a torus. 
In Ref. \onlinecite{Lee05} the Laughlin state is studied numerically and the properties of the rapid crossover from the 
inhomogeneous TT-state to a virtually homogeneous state are determined in detail. Moreover, the gap is confirmed to be non-vanishing for all radii in 
accordance with the general argument referred to above.

\section{Generalizations}

The discussion above and the work referred to, with the exception of that by Su \cite{su}, concerns the Laughlin filling factors $1/(2k+1)$ only. 
For other filling factors the situation is as follows. The interacting electron gas in a single 
Landau level can be exactly diagonalized  in the TT-limit for any rational filling factor $p/q$  \cite{bk3,bk2}. This holds provided the interaction between the electrons, which is purely electrostatic in this limit,  obeys the concavity condition $V''_k \equiv V_{k-1}+V_{k+1}-2V_{k}>0$, where $V_k$ is the electrostatic interacting energy between two electrons $k$ lattice constants apart. This is a very generic condition obeyed by, for example, the Coulomb and short range interactions. The ground state, 
which is $q$-fold degenerate,  is a single Slater determinant where the electrons are as far apart as possible in a unique and well-defined sense, there is a gap to excitations 
and there are quasiparticles with charge $\pm e/q$---these are domain walls between the degenerate ground states.  
For example, the ground states at $2/5$ and $4/11$ have unit cells $00101$ and $00100100101$ respectively. According to the above, $01$ is the
$-e/3$ quasielectron in the $1/3$ ground state, hence the $2/5$ and $4/11$ ground states can be viewed as condensates of these quasielectrons in the $1/3$ ground state.
This generalizes to any $p/q$: the ground state  is a condensate of quasiparticles in a parent state---for odd $q$ this is the original  
Haldane-Halperin hierarchy construction \cite{hierarchyHLH}, 
which can thus be derived in the TT-limit. Moreover, the gap is in the TT-limit a monotonously 
decreasing function of the denominator $q$, leading to the prediction that, in a given sample, abelian QH states should be observed for all 
$p/q$, with $q \le q_0$, for some $q_0$ \cite{hierarchy,bk3}---in excellent agreement with the highest mobility data of Pan et al \cite{pan}. 
The TT-ground states can also be interpreted as being obtained by filling effective Landau levels \cite{bk3}, in accordance with the composite fermion picture \cite{jain}. Thus, in 
the TT-limit, it is clear that the old hierarchy and composite fermions are two ways of viewing the same thing rather than being fundamentally different. 
Recently the exact equivalence between the states in the positive Jain sequence and a set of hierarchical wave functions has been proven also in the bulk case  \cite{hans}. Furthermore, assuming that phase transitions occur only by condensation of the charge $\pm e/q$ quasiparticles, the global phase diagram \cite{global,lutken} is obtained in 
the TT-limit \cite{hierarchy}.

Moving away from the TT-limit, exact diagonalization of small systems shows, for Coulomb interaction in the lowest Landau level,  a transition to a new state  if the denominator $q$ is even, and no transition if $q$ is odd. The qualitative properties 
of the TT-states agree with those of abelian QH states and the TT-states are the limits of Jain's composite fermion wave functions \cite{jain}, for filling factors where these exist. We conclude that
the TT-states, for odd $q$, are adiabatically connected to abelian QH states just as is the case for the Laughlin fractions. Further support for this is given by the 
conformal construction of hierarchy states \cite{cft}. Of course, an abelian QH state need not be 
observed at a particular filling factor---in such a case a phase transition occurs as one adiabatically continues from the TT-limit to the physical interaction.

For $\nu=1/2$, a phase transition from the gapped TT-state into a Luttinger liquid occurs when the radius is of the order one magnetic length \cite{bk1}. Exact diagonalization studies \cite{bk2}
indicate  that this gapless system develops continuously into the composite fermion wave function \cite{rr} believed to describe the observed metallic state in the half-filled lowest 
Landau level. The transitions observed in exact diagonalization for other even $q$, when the radius grows, indicate that a similar scenario may hold 
for other even denominator fractions.  

We end this Section with references to related work. A simple representation of the non-abelian Moore-Read state \cite{mr} can be obtained in the TT-limit \cite{nonabelianTTwe, nonabelianTTthey}. The quasiparticles 
are again domain walls between the degenerate ground states and the correct non-trivial degeneracies are reproduced. This has been generalized 
to other states \cite{other} and is related to the approach by Bernevig and Haldane \cite{bh}.\footnote{The TT-states are the root partitions of the Jack polynomials in Ref. \onlinecite{bh}.}
There is also a close relation to the recent work by Wen and collaborators \cite{wen}. The periodicity of 
the Laughlin wave function on the thin cylinder has been rigorously proven by Jansen et al \cite{Jansen}.  The Halperin $(mm'n)$ states are discussed by Seidel and Yang \cite{SeidelYang}.

\section{Fractional Statistics}
The quasiparticles in the quantum Hall effect not only have fractional charge but they also obey fractional exchange statistics \cite{jml, frac}, as well as exhibit exclusion 
statistics \cite{exclusion}. Since both types of statistics are fractional, we refer to the former as exchange statistics to make the distinction clear.  In this section we discuss to what extent these features exist in the TT-limit. As a prelude we show that the correct  Aharonov-Bohm phase factor is obtained. 

\subsubsection{The Aharonov-Bohm effect}
We impose periodic boundary conditions so that the cylinder becomes a torus and insert a magnetic flux, $\Phi$, through the hole that is formed in the process. Consider, to be definite, a quasihole at $\nu=1/3$. This consists of an extra zero inserted somewhere in the string of unit cells 001 making up the ground state. This quasihole is moved around the torus by consequtively moving the electron to the right of the hole one step to the left. When the hole has completed three full turns around the torus, one electron has effectively moved the full length of the torus and, assuming standard minimal coupling of the electron to the vector potential, it has picked up the phase $e^{i e\Phi/\hbar}$, and thus it is natural to assign the phase $e^{i e\Phi/3\hbar}$ to the process of moving the hole one turn---this is the appropriate Aharonov-Bohm phase for a charge $e/3$ particle.\footnote{Although the phase corresponding to three turns is unambiguous, there is a gauge ambiguity in as
 signing the phase for a single turn.}
\subsubsection{Exchange statistics}  

The calculation of the exchange statistics for Laughlin quasiholes  is on  a firm basis, because the powerful plasma analogy introduced by Laughlin can be used to analytically evaluate the Berry phases \cite{arovas}. 
This is not the case for the quasiparticles in most other QH states (including the quasielectrons in the Laughlin states), where the arguments for exchange statistics are based on a combination of heuristic reasoning based on effective Chern-Simons theories, arguments based on monodromies of conformal blocks \cite{read}, and numerics \cite{hansson09}. It would thus be of great interest if a convincing argument for exchange statistics could be given for QH states in the TT-limit where a simple generalization from Laughlin states to general hierarchical states, and even to non-abelian states, could be hoped for. 

Exchange statistics is defined as follows. Using single-valued wave functions, the fractional statistics phase is obtained by adiabatically interchanging two quasiparticles and calculating the Berry phase.
For this to be possible the quasiparticles must be localized, but the ones in the TT-limit are domain walls extending all around the system in
one direction. One might think that it is possible to form wave packets localized in two dimensions by superposing domain walls with different positions: $\Psi(\{{\bf r}_i \})=\sum_\alpha a_\alpha \Psi_\alpha(\{{\bf r}_i \})$, where $\Psi_\alpha$ is the state with a quasihole in unit cell  $\alpha$. The problem with this approach becomes apparent when calculating the density of the electron gas $\rho({\bf r})=\int\prod_{i=2}^{N_e}d^2{\bf r}_i |\Psi (\{{\bf r}_j \})|^2$. Since $\Psi_\alpha$ is a single Slater determinant of one-electron states, $\psi_k({\bf r}) \propto e^{2\pi i kx/L}$ (where $x$ is the coordinate around the cylinder), it follows that 
$\int\prod_{i=2}^{N_e}d^2{\bf r}_i \Psi^*_\alpha \Psi_\beta $ is non-zero only if $\Psi_\alpha$ and $\Psi_\beta$ differ by at most a single one-electron state; this is the case only if $|\alpha-\beta|=0,1$, and then the  one-particle states that do not match will have momenta $k$ that differ by one.  As a consequence, $\rho ({\bf r})$ contains only the modes $1, e^{\pm 2\pi i x/L}$, and a hole localized in the $x$-direction cannot be constructed. 
Thus we conclude that a Berry phase calculation of exchange statistics cannot be performed in the TT-limit. One may be tempted to allow 
states that are more general than the pure TT-domain walls in the superposition in order to build a localized quasiparticle. 
However, to apply adiabatic reasoning, one must make sure that the localized quasiparticle state is an approximate\footnote{
To be precise, one must imagine adding a weak external potential which localizes the quasiparticles and eliminates the zero mode due to translation---this is what is always tacitly assumed in calculations of Berry phases, where a finite gap to excited states is mandatory. }
eigenstate to the Hamiltonian in the two-quasiparticle sector. Thus, adding other components to the quasiparticle wave function in order to localize it one must at the same time change the Hamiltonian by modifying the static interaction of the TT-limit and also to include hopping terms. 
We do not see any way of achieving this,  barring using the known quasiparticle excitations in the Laughlin ground state that is explicitly known for any radius. 

Recently, Seidel and Lee calculated the exchange statistics on a torus using that the Laughlin quasihole state is connected to the TT-state by an adiabatic change of the aspect ratio \cite{SeidelLee}. This amounts to a novel way of extracting the fractional exchange statistics of the Laughlin quasiholes without using the plasma analogy, but it  does use the specific form of the Laughlin wave function so the generalization to hierarchy states requires additional assumptions. This method has been applied also to the non-abelian exchange statistics in the Moore-Read state \cite{Seidel08}.

\subsubsection{Exclusion statistics}
\begin{figure}[h!]
\begin{center}
\resizebox{!}{20mm}{\includegraphics{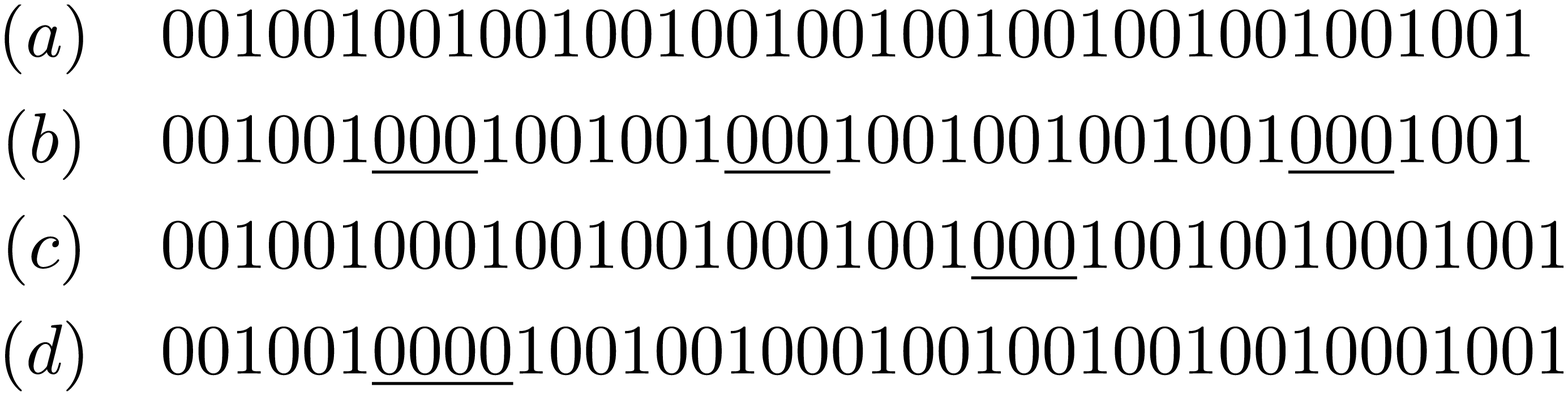}}
\end{center}\caption{\textit{{\small Exclusion statistics for quasiholes. 
(a) The TT-ground state at $\nu=1/3$ with $N=13$ unit cells.
(b)  Three quasiholes (underlined) created in the state in (a), and one unit cell removed to keep the size of the system unchanged. 
(c) One additional quasihole (underlined) created in a unit cell that did not already contain a quasihole. 
(d) One additional quasihole (underlined) created in a unit cell that  already contained  a quasihole. The $N-1$ states in (c) and (d) are the low energy quasihole states.
}}}\label{stat1}
\end{figure}

We now turn to exclusion statistics; this 
can be determined by simple counting in the TT-limit. As an example, we consider $\nu=1/3$ where quasiholes are created by inserting zeroes in the state with unit cell 001, see Fig.\ref{stat1}.  Assume the system initially contains $N$ electrons on a torus, {\it ie} periodic boundary conditions are imposed. Following Haldane's definition \cite{exclusion}, one first inserts three separated zeroes and removes one unit cell to keep the size of the system unchanged. One then counts the number of low energy states available for an extra zero, assuming the original zeroes have fixed positions. 
In addition to $V''_k  \equiv V_{k-1}+V_{k+1}-2V_{k}>0$, which implies that the TT-state is stable,  we assume that $V''_k /V''_{k+1}\rightarrow \infty$ in the TT-limit---this allows us to include only the leading terms when calculating the energies of the excitations.\footnote{This condition holds 
for an exponentially decaying interaction, but for a longer-range interaction, such as Coulomb, only the weaker condition $V''_k > V''_{k+1}$ holds. In the latter case further range terms must be considered and may, in principle, affect the exclusion statistics.} 
The minimal neutral excitation energy in the ground state at 1/3 corresponds to one electron moving one lattice constant; this excitation has energy $V''_3$.
Obviously, there are many excitations at, and above, this energy scale---both in the ground state and in a state with quasiparticles present. The low energy states available for the extra hole consists of the excitations below this "neutral gap". The hole can be inserted anywhere between the electrons, leading to the $N-1$ different states in Fig.\ref{stat1}b and c. These states are the lowest energy states.  They are the only states that have the three initial quasiholes in fixed positions and do not contain a sequence 11 or 101.\footnote{The position of the  three initial quasiholes is fixed up to the translation implied by the insertion of the fourth quasihole.} Their differences in energy are of order $V''_4$ at most. The higher energy states that have lowest energy are obtained by moving one electron to a nearby site so that a sequence 101 is formed;  this costs energy $V''_3$ at least.  In the ground state in Fig.\ref{stat1}a, a low energy quasihole can be added in $N$ 
 places and, after adding three quasiholes and removing one unit cell, it can be added in $N-1$ places. Thus the number of available states for a quasihole has decreased by 1/3 for each added quasihole, hence  the exclusion statistics parameter is $g_{--}=1/3$ in agreement with Haldane's claim. ($-g_{\alpha \beta}$ is the change in the number of available states for a quasiparticle of type $\alpha$ per quasiparticle of type $\beta$ that has been added.)

\begin{figure}[h!]
\begin{center}
\resizebox{!}{20mm}{\includegraphics{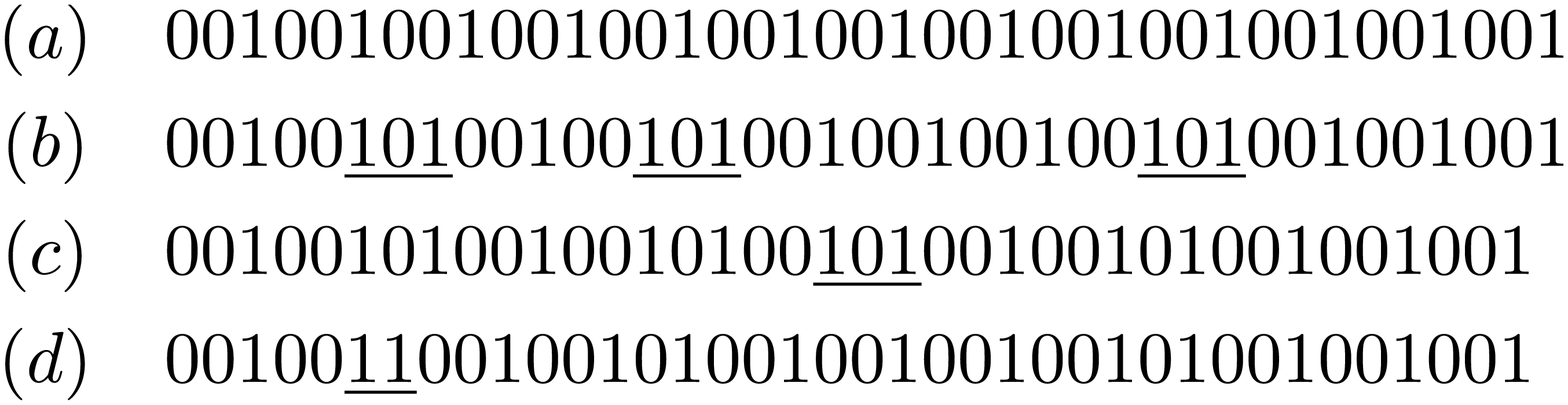}}
\end{center}\caption{\textit{{\small  Exclusion statistics for quasielectrons. 
(a) The TT-ground state at $\nu=1/3$ with $N=13$ unit cells.
(b)  Three quasielectrons (underlined) created in the state in (a), and one unit cell added to keep the size of the system unchanged. 
(c) One additional quasielectron (underlined) created in a unit cell that did not already contain a quasielectron. 
(d) One additional quasielectron (underlined) created in a unit cell that already contained a quasielectron. 
Only the $N-2$ states in (c) are the low energy quasielectron states.
}}}\label{stat2}
\end{figure}

Repeating the argument just given for quasielectrons, which are created by removing zeroes, one finds something interesting, see Fig.\ref{stat2}. First one removes three separate zeroes and adds one unit cell. Removing the next zero leads to $N+1$ different states and one might expect that the statistics parameter is $-1/3$ in accordance with Haldane's original suggestion. However, the three states in Fig.\ref{stat2}d, which contain two electrons next to each other have energy of order $V''_2$ above the others. Only the $N-2$ states in Fig.\ref{stat2}c are low energy states. They are the only states, with the three quasielectrons in fixed positions, that do not contain a sequence 11, and their differences in energy are of order $V''_4$ at most. The higher energy states that have lowest energy are obtained by moving one electron to a nearby site so that a sequence 11 is formed;  this costs energy $V''_3$ at least. In the ground state in Fig.\ref{stat2}a, a low energy quasielec
 tron can be added in $N$ places and in Fig.\ref{stat2}b, after adding three quasielectrons and removing one unit cell, it can be added in $N-2$ places. Thus the number of available states for a quasielectron has decreased by 2/3 for each added quasihole, hence $g_{++}=2/3$. That also this statistics parameter is positive is physically sensible, since a negative coefficient would imply that the number of single quasielectron states would increase with the number of quasielectrons present.\footnote{The asymmetry between quasielectrons and quasiholes was noticed by Canright and Johnson \cite{canright}. However, based on exact diagonalization of small systems on the sphere these authors obtained $g_{++}=5/3$. This discrepancy may be due to the difference in geometry or interaction.}

\begin{acknowledgments}
We would like to thank Eddy Ardonne, Emil J. Bergholtz and Maria Hermanns for useful discussions.
This work was supported by the Swedish Research Council and by NordForsk.
\end{acknowledgments}

\end{document}